# Measuring Understanding Through Discrete Compositional Knowledge Structures in Hierarchical Automata

Igor Balaz[1][0000-0002-6831-9232]

[1] Neovivum Technologies, Serbia
Igor.balaz@neovivum.com

**Abstract.** How do we measure genuine understanding in artificial cognitive systems? Current approaches face a measurement gap: probabilistic systems refine confidence gradually, practice-based systems compile knowledge through repeated execution, and neural systems distribute understanding across opaque embedding spaces.

We propose that making understanding measurable requires architectures where understanding formation produces discrete, inspectable structural signatures. This paper presents hierarchical automata built from finite state machines representing patterns and higher-order automata representing compositions. Constraint inference constructs automata from single observations. Similarity detection clusters related automata, making concept robustness quantifiable. Graph memory makes compositional knowledge directly inspectable. Metacognitive mechanisms enable observable reconfiguration.

We demonstrate understanding measurement in a simple geometric domain. Graph evolution tracking reveals five measurable signatures: immediate representation formation, structural knowledge, generalization capacity, compositional awareness, and metacognitive access. These measurements distinguish structural understanding from statistical correlation.

Our contribution is a framework for making understanding measurable through discrete compositional knowledge structures. This measurement capability complements perceptual learning in neural systems and task execution in neurosymbolic architectures.

## 1 Introduction

The question of how to measure genuine understanding in cognitive artificial systems has become central to AGI research. Large language models demonstrate remarkable correlational capabilities (Vaswani et al., 2017), while cognitive architectures refine knowledge through iterative inference (Laird, 2012) or practice-based compilation (Anderson, 2007). But when we ask whether these systems genuinely understand what they process, we face a measurement problem rooted in architectural constraints. Systems that learn through statistical correlation cannot distinguish structural comprehension from pattern matching at scale, and systems that develop understanding through practice distribute learning across episodes. Understanding remains either opaque within distributed representations or emerges gradually through accumulation, offering no discrete signature we can track or verify. The question "does it understand?" collapses into "has it seen enough examples?"—a quantitative threshold rather than a qualitative difference.

We propose that making understanding measurable requires an architecture where understanding formation produces discrete, inspectable structural signatures. This paper presents a minimal cognitive system built on hierarchical automata using finite state machines (FSMs) and higher-order automata (HOAs) that demonstrate three measurable properties absent in existing approaches: immediate representation formation from single exposure, explicit compositional dependencies that reveal knowledge structure, and metacognitive reconfiguration that demonstrates awareness of what the system knows. These properties emerge from core architectural mechanisms. Constraint inference produces discrete representations from structural observation rather than statistical accumulation. Hierarchical graph memory makes compositional dependencies explicit through traversal. Dependency detection enables metacognitive awareness when components change or disappear.

None of the existing approaches can answer "when did it understand?" with a discrete event or "what does it understand structurally?" with an inspectable representation; post-hoc explainability methods (Ribeiro et al., 2016; Lundberg & Lee, 2017) approximate decision explanations but do not make understanding itself structurally inspectable. Our main contribution is a measurement framework: the five criteria are derived from philosophy of mind and cognitive science, and the architecture is designed to produce measurable instances of them. We do not claim this constitutes understanding in a deep philosophical sense. Instead, we demonstrate what structural signatures of understanding would look like and how they can be measured. Where existing approaches offer impressive capabilities but limited measurement access, hierarchical automata make understanding formation observable as discrete events, compositional structure inspectable through graph traversal, and concept robustness quantifiable through topology metrics.

The paper proceeds as follows. Section 2 examines existing approaches and their measurement gaps. Section 3 establishes theoretical criteria for measurable understanding. Section 4 describes the architecture. Section 5 demonstrates understanding formation through the "house" example with quantitative metrics. Section 6 analyzes how these measurements differ from existing approaches and what they reveal about structural versus statistical understanding. Section 7 discusses implications for AGI research, connections to human cognition, and limitations.

## 2 Related work

This section examines why current architectures cannot provide discrete, inspectable signatures of understanding formation.

Cognitive architectures represent the most sustained effort to build general intelligence through explicit cognitive mechanisms, but they require iterative refinement or extended practice to form understanding. OpenCog (Goertzel et al., 2014) constructs a weighted hypergraph where understanding emerges through iterative confidence updates as the Probabilistic Logic Networks subsystem propagates truth values through repeated inference. SOAR (Laird, 2012; Laird et al., 1987) approaches understanding through goal-driven problem-solving, where the chunking mechanism creates production rules only after the system encounters and resolves impasse states. Crucially, SOAR does not learn from pure observation—presenting a pattern without an associated goal or problem-solving context produces no impasse and therefore no learning. ACT-R (Anderson, 2007; Anderson & Lebiere, 1998) separates declarative chunks from procedural productions and relies on repeated retrieval and execution to strengthen knowledge through production compilation. CLARION (Sun, 2016) extracts symbolic rules from implicit neural patterns through gradual refinement, while Sigma (Rosenbloom, 2013; Rosenbloom et al., 2013) implements all cognitive functions as probabilistic inference over factor graphs, requiring multiple observations to estimate distributions and build confidence in relational patterns. These architectures share a fundamental measurement constraint: understanding formation happens through accumulation of statistical evidence or compilation through practice. We cannot identify when understanding formed because the transition occurs through continuous refinement rather than discrete recognition.

Neurosymbolic architectures combine neural perception with symbolic reasoning and make composition architecturally explicit. Neural Module Networks (Andreas et al., 2016) assemble task-specific networks from reusable modules, while DeepProbLog (Manhaeve et al., 2021) integrates neural predicates with probabilistic logic programs. These approaches achieve strong performance on compositional reasoning tasks, but they measure task execution, not understanding formation. We cannot observe when the system learns that a concept should decompose into specific modules or when predicates should combine in particular logical patterns. The compositional structure exists to solve tasks, and understanding remains implicit in learned module weights and hand-specified logic programs rather than explicit in evolving graph topology.

Few-shot learning methods achieve rapid adaptation to new tasks with minimal examples at test time. MAML (Finn et al., 2017) meta-learns an initialization that enables fast fine-tuning through gradient descent, while Prototypical Networks (Snell et al., 2017) learn metric spaces where classification reduces to nearest-neighbor comparison. But this sample efficiency requires extensive meta-training across hundreds of tasks to discover good initializations or thousands of episodes to learn discriminative embeddings. The "few-shot" capability refers to test-time adaptation after this meta-training phase. We cannot measure when these systems form understanding because the meta-training process distributes learning across many tasks, and understanding emerges from statistical patterns over the entire task distribution rather than from structural analysis of individual observations.

Transformer architectures (Vaswani et al., 2017) process sequences by learning attention weights that relate tokens to semantic context, capturing statistical regularities where words that co-occur in similar contexts acquire similar embeddings. But these representations offer no access to compositional structure. When a language model processes "the house has walls and a roof", the embedding vectors capture statistical relationships, but we cannot inspect the model to determine whether it represents the structural dependency that houses require walls and roofs. The knowledge is distributed across millions of parameters in a high-dimensional space optimized for prediction, not structural comprehension. Recent work on mechanistic interpretability (Elhage et al., 2021) attempts to identify circuits within transformer models, but this analysis remains post-hoc and partial—we can observe attention patterns but cannot extract a compositional graph showing what concepts depend on what components. Understanding, if it exists, remains opaque.

The next section shows how hierarchical automata address this gap by making understanding formation discrete, compositional structure explicit, and metacognitive awareness observable.

## 3 Theoretical Framework: What Makes Understanding Measurable?

The question "how do we measure understanding?" requires first answering "what is understanding?" This section establishes five criteria drawn from philosophy of mind and cognitive science, then shows how each criterion maps to a measurable computational property in hierarchical automata.

**Structural Knowledge:** A system that genuinely understands knows not just what patterns exist but why they exist. Statistical correlation shows roofs and houses co-occur frequently; structural understanding shows houses require roofs because the compositional relationship is mandatory, not optional. This distinction appears throughout philosophy of mind: Searle's Chinese Room argument (Searle, 1980) challenges symbol manipulation without meaning, while Harnad's symbol grounding problem (Harnad, 1990) asks how symbols acquire semantic content beyond associations. Understanding requires knowing structural necessity rather than statistical regularity.

Measuring this requires access to causal or compositional relationships. A system demonstrates structural knowledge when we can inspect its representation and extract dependencies: 'X requires Y' rather than 'X correlates with Y (with probability p)'. The representation must make necessity explicit, not just co-occurrence frequency.

**Compositional Awareness:** Human understanding is compositional. We understand "house" in terms of components (walls, roof, doors, windows) and understand how these components compose into the whole. This compositional structure enables both generalization (a cottage and a mansion are both houses despite differences) and decomposition (we can reason about individual components). Fodor's language of thought hypothesis (Fodor, 1975) posits that concepts have combinatorial structure, and developmental psychology demonstrates that children learn categories by understanding part-whole relationships, not just surface features (Mandler, 2004).

Measuring compositional awareness requires inspecting knowledge architecture. A system demonstrates this property when its representation reveals hierarchical structure: what contains what, what is built from what.

**Generalization Capacity:** Understanding enables generalization beyond specific instances. Seeing one house allows humans to recognize other instances, even with significant variation. This differs from rote memorization, which stores instances without abstracting commonalities. Rosch's work on prototypes and basic-level categories (Rosch, 1978) shows that human concepts organize around central examples with graded membership, not rigid definitions. Generalization requires forming abstract representations that capture invariances across variations.

Measuring generalization requires tracking how representations handle variation. A system demonstrates this capacity when we can observe cluster formation where similar instances group together, creating robust concepts that survive variation.

**Metacognitive Access:** Understanding includes knowing what you understand. Metacognition is awareness of one's own knowledge state, including recognizing gaps and monitoring comprehension. Flavell's work on metacognitive monitoring (Flavell, 1979) demonstrates that effective learners track their understanding and adjust strategies when gaps appear. This self-awareness distinguishes genuine understanding from blind pattern matching.

Measuring metacognition requires observable responses to knowledge perturbations. A system demonstrates metacognitive access when structural changes (component deletion or discovery of deeper primitives) trigger reorganization that reflects awareness of dependencies. We use 'metacognitive access' to denote this functional property (dependency-aware reconfiguration triggered by structural change), but we acknowledge that whether this constitutes metacognition in the full cognitive-science sense remains an open question.

**Sample Efficiency:** Human understanding often requires minimal exposure. We can see one example of a new animal category and recognize other instances, even with variation. This sample efficiency distinguishes human learning from systems requiring thousands of examples to extract statistical regularities. Carey and Bartlett's work on fast mapping (Carey & Bartlett, 1978) shows that children acquire word meanings from single exposures, refining these initial representations through experience. So, understanding formation happens quickly while concept robustness develops gradually.

Measuring sample efficiency requires tracking representation formation relative to exposure count. A system demonstrates this property when we can observe immediate representation creation from single instances, distinct from gradual concept refinement through multiple instances.

The next section describes our architecture that implements mechanisms for measuring these criteria: FSM pattern representation, HOA compositional structure, constraint inference for single-exposure learning, similarity-based clustering, and mapping back to the five criteria.

## 4 Architecture

Our architecture implements the outlined theoretical framework through three automata types organized in a graph memory structure. Finite state machines (FSMs) represent basic patterns, higher-order automata (HOAs) represent compositional structures, and the learning engine constructs both through constraint inference from observation. The components are domain-agnostic by design. FSMs represent any pattern definable through states and constrained transitions. These are not only geometric shapes but also grammatical patterns (states as syntactic roles, transitions as dependency rules) or musical phrases (states as note positions, transitions as interval constraints). HOAs then represent any compositional structure built from such patterns. The geometric domain is used throughout for clarity of illustration, not as a domain restriction. This section describes each component and shows how they produce measurable understanding signatures.

**Finite State Machines (FSM):** An FSM provides pattern representation as a directed graph with states as nodes and transitions as edges. Each transition carries a constraint specifying the input condition that triggers the state change and any spatial or temporal requirements that must hold. The learning engine constructs an FSM by observing a pattern and inferring what states and transitions are necessary to capture its structure.

For instance, if a square is displayed, the learning engine observes four sequential line segments with specific spatial constraints: each segment connects to the next at 90-degree angles, and the fourth segment connects back to the first. The resulting FSM contains four states (one per side) with transitions labeled by geometric constraints.

Such representation enables direct inspection of what inputs trigger it and what spatio-temporal relationships must hold.

**Higher-Order Automata (HOA):** An HOA represents composition by organizing component automata in a hierarchical graph. The HOA graph contains FSMs and other HOAs as subgraphs, connected by compositional edges that specify containment and dependency relationships.

For instance, a house HOA contains FSMs for walls (square FSM) and roof (triangle FSM) as component subgraphs. Compositional edges specify structural constraints: walls are the foundation, and the roof sits atop walls.

Such hierarchical organization enables compositional inspection. Traversing from the house node through compositional edges reveals wall and roof, each with its own internal structure.

**Learning Engine:** The learning engine constructs automata through constraint inference from pattern observation. When presented with a novel pattern, it analyzes the structure to identify necessary states, required transitions, and mandatory constraints. The inference mechanism operates through structural decomposition. A pattern is analyzed into constituent elements, temporal ordering if applicable, and compositional hierarchy. For each element, the engine determines what constraints must hold. For instance, a square requires four segments of equal length, perpendicular connections, and closure. The construction is immediate. If the observed pattern contains subpatterns matching existing automata in graph memory, the engine constructs an HOA referencing these components. Otherwise, it constructs an FSM treating the pattern as monolithic. This creates order-dependent learning: observing rectangles before houses produces house HOAs, while observing houses first produces house FSMs.

When analyzing subsequent observations of similar patterns, the engine distinguishes dimensional variations from structural variations. Patterns differing only in parametric values (dimensions, proportions) match existing FSM structure and require no new representation. Patterns with structural differences (altered state sequences, different constraint types) create new FSMs that cluster through similarity detection.

**Similarity Detection:** The similarity detection mechanism compares automata to identify structural overlap. When two automata share significant constraint structure but differ in details, the mechanism creates a similarity edge weighted by the degree of overlap. Similarity is computed by comparing constraint graphs. Consider two square FSMs: one axis-aligned, one rotated 45 degrees. These share core constraint structure but differ in orientation. The similarity detector creates an edge connecting them, weighted by structural overlap. As our system encounters variations, similarity edges accumulate. Multiple rotated squares, stretched rectangles, and trapezoids connect to the original square FSM with varying edge weights. The resulting cluster structure enables concept clustering, providing measurable concept robustness through average path length, connection degree distribution, and modularity metrics.

**Graph Memory:** The graph memory implements the knowledge architecture as a directed graph stored in a graph database. Nodes represent automata (FSMs or HOAs), and edges represent relationships (similarity, composition, dependency).

This graph structure makes knowledge inspectable. We can query which automata exist, how they relate, and what constraints each contains. Graph evolution is trackable: every learning episode adds nodes, creates edges, and constructs hierarchies. These modifications are logged with timestamps.

**Metacognitive Mechanisms:** Our system accesses its own knowledge structure by inspecting the graph. This external inspection capability enables metacognitive awareness of what concepts it knows, how concepts relate, and what dependencies exist.

The system exhibits metacognitive behavior through two types of reorganization. When a component automaton is deleted, our system detects broken dependencies by identifying HOAs that contained the deleted component. These HOAs then reconfigure by finding alternative components, restructuring composition, or marking themselves incomplete. When deeper primitives are discovered, the system restructures existing representations by converting FSMs to HOAs and adjusting compositional hierarchies.

Such reconfiguration behavior provides measurable metacognitive signatures. For instance, deleting a wall FSM reveals which house HOAs respond, how they reorganize, and whether they maintain structural coherence. Introducing line primitives reveals how the system restructures rectangle and triangle representations into hierarchical compositions.

In the outlined architecture each component produces following measurable signatures: (i) FSMs encode constraints explicitly; (ii) HOAs encode composition hierarchically; (iii) the learning engine creates discrete representations; (iv) similarity detection creates quantifiable clusters; and (v) graph memory makes knowledge architecture inspectable.

**Table 1.** From Architecture to Theoretical Criteria.

| Understanding Criterion | Computational Mechanism | Measurable Output |
|---|---|---|
| Structural Knowledge | Constraint graphs in FSMs | Inspectable dependencies |
| Compositional Awareness | Hierarchical HOA structure | Traversable composition |
| Generalization Capacity | Similarity-based clustering | Quantifiable topology |
| Metacognitive Access | Graph inspection + reconfiguration | Observable reorganization |
| Sample Efficiency | Immediate automaton construction | Discrete node addition events |

The architecture described in Section 4 is fully implemented in Python, with a Neo4j graph database serving as the graph memory backend. All metrics reported in the next section were computed by running the implemented system on the geometric stimulus sequences described below.

# 5 Demonstration

This section demonstrates measurable understanding formation using a simple geometric domain where primitives (rectangles, triangles) compose into structures (houses). Such seemingly trivial simplicity isolates the measurement question from domain complexity. To show how these measurements reveal understanding formation and refinement, we track graph evolution through four phases, and extract quantitative metrics.

**Experimental Setup**

**Domain**: The geometric domain contains two abstraction levels. Primitives include rectangles (four perpendicular sides with parallel opposites) and triangles (three connected segments forming closure). Composites include houses combining rectangles as walls and triangles as roofs with spatial relationships.

**Controlled Exposure Sequence**: Primitives are presented before composites so houses can form as HOAs referencing component automata rather than monolithic FSMs. This reflects realistic compositional learning. Phase 1 presents primitives to measure immediate representation. Phase 2 presents variations that either match existing structures (dimensional changes) or create new FSMs (structural modifications), measuring concept refinement through clustering. Phase 3 presents composite structures to measure hierarchical understanding. Phase 4 demonstrates metacognitive reconfiguration through component deletion and hierarchical restructuring.

**Graph Evolution Tracking**: The graph database logs all modifications with timestamps. Each automaton addition records when it was created, what pattern it represents, and what constraints it contains. Each similarity edge records when it was created, which automata it connects, and what overlap weight it carries.

**Measurement Protocol**: At each phase, we extract graph metrics: node and edge counts, cluster topology (path length, modularity), compositional depth, and reconfiguration patterns.

**Phase 1: Immediate Representation Formation**

**Input:** The system observes a single scene containing one rectangle and one triangle in spatially separated positions.

**Learning Process:** The learning engine analyzes the scene through structural decomposition. It identifies two distinct geometric patterns, each occupying separate regions. For the rectangle, it infers perpendicular-sides constraints and parallel-opposites constraints. For the triangle, it infers three-segment closure constraints. The engine constructs an FSM for each pattern encoding these structural properties.

**Graph Evolution and Measurements:** The system adds two nodes: one rectangle FSM and one triangle FSM. The modification happens in a single transaction, timestamped to mark when both primitive representations were created. The graph evolution log shows discrete node addition with precise constraint structure, confirming discrete learning events with one-to-one exposure-to-representation mapping.

**Phase 2: Primitive Concept Refinement Through Clustering**

**Input:** The system observes five rectangle variations and five triangle variations. Rectangle variations include three with only dimensional differences (aspect ratios 1:2, 1:4, and 1:1), one with rounded corners, and one rotated 45°. Triangle variations include three with only dimensional differences (equilateral, isosceles, scalene), one inverted (apex down), and one with curved sides.

**Learning Process:** The learning engine analyzes each variation. Variations differing only in dimensional parameters (width, height, proportions, angles) match the structure of existing FSMs since the core constraint pattern remains identical, requiring no new FSM. Structural modifications create new FSMs: the rounded-corner rectangle introduces curved transitions, while the rotated rectangle produces a different state-traversal sequence. The same analysis applies to triangles, where dimensional variations match the existing FSM while structural modifications (inversion, curved sides) create new FSMs. The similarity detection mechanism compares all FSMs based on constraint overlap, creating weighted similarity edges for both rectangle and triangle clusters.

**Graph Evolution and Measurements:** The system adds four new FSMs: two for structurally distinct rectangles and two for structurally distinct triangles. The resulting structure comprises two clusters: three rectangle FSMs with internal connectivity and three triangle FSMs with internal connectivity. Cluster topology quantifies concept robustness. Average path length between nodes in each cluster is 1.0 (fully connected). Modularity is 0.92, indicating strong cluster separation. Connection degree distribution emerges from similarity patterns, where FSMs closer to cluster average have higher connectivity than structurally peripheral examples. This demonstrates that similarity-based clustering creates measurable concept structure at the primitive level while constraint-based inference prevents redundant FSM creation for purely dimensional variations.

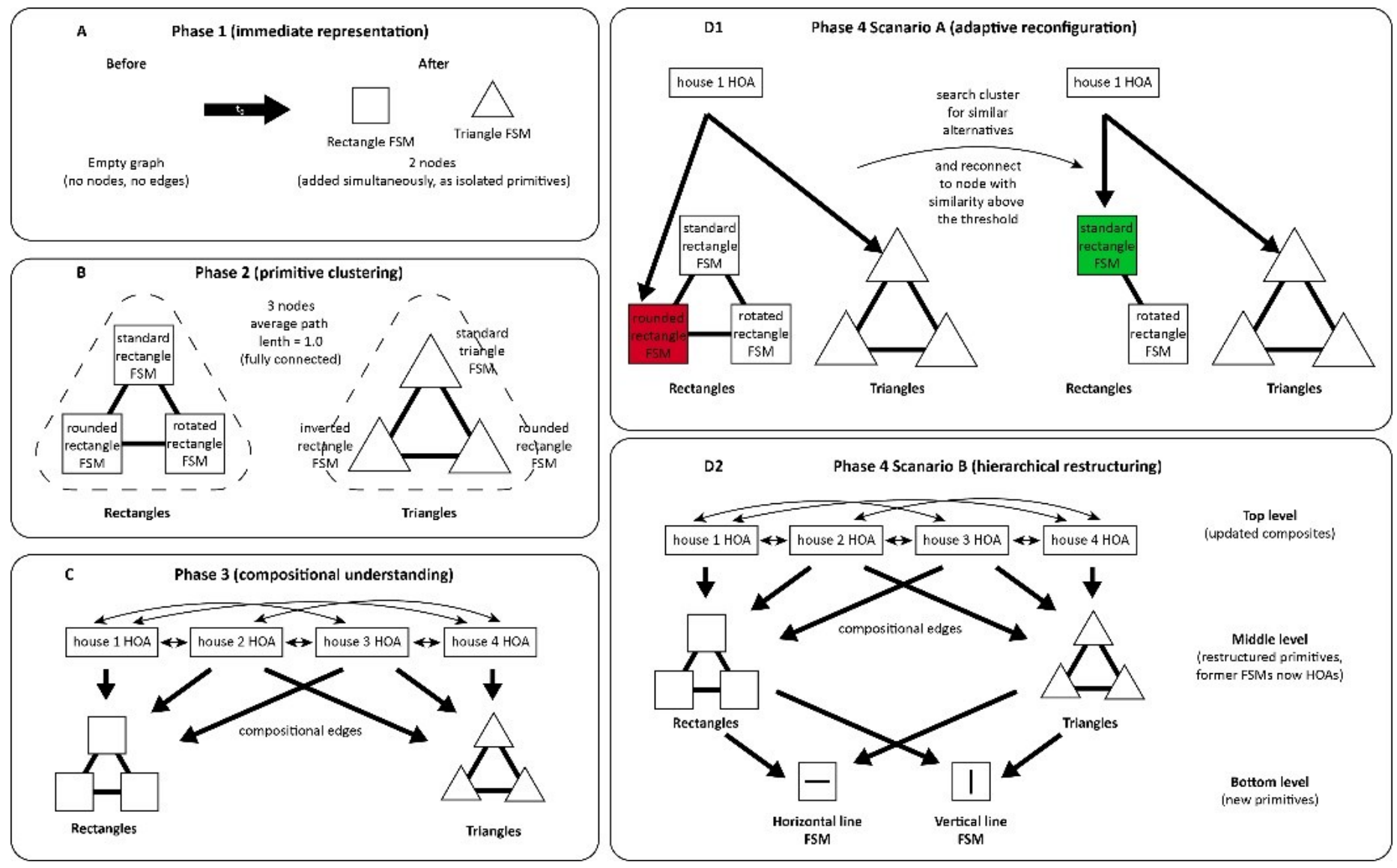

**Fig. 1.** visualizes graph evolution across all four phases. Panel A shows discrete node addition in Phase 1. Panel B shows primitive clustering in Phase 2. Panel C shows compositional construction and clustering in Phase 3. Panel D1 shows component deletion in Phase 4 Scenario A. Panel D2 shows hierarchical restructuring in Phase 4 Scenario B.

### Phase 3: Compositional Understanding and Clustering

**Input:** The system observes four house patterns. Each house contains a rectangle (vertical, serving as wall) with a triangle positioned atop (serving as roof). The four houses vary in wall height, roof steepness, and overall proportions.

**Learning Process:** For the first house, the learning engine detects that the pattern contains subpatterns matching existing automata in graph memory (rectangle FSM and triangle FSM). It constructs a house HOA with compositional edges to these component FSMs, encoding spatial constraints where rectangle is vertical (wall) and triangle sits atop rectangle (roof). For subsequent house variations, the same compositional construction occurs, followed by similarity detection.

**Graph Evolution and Measurements:** The system adds four house HOA nodes with compositional edges to rectangle and triangle FSMs from prior phases. The similarity detector creates edges connecting the house HOAs based on shared compositional structure despite variation. The resulting house cluster demonstrates that clustering mechanisms operate at compositional levels. Measurements confirm compositional depth of 1, dependency count of 2 components per house, and cluster connectivity (average path length 1.3). This demonstrates that compositional construction and similarity-based clustering produce measurable signatures at hierarchical levels.

### Phase 4: Metacognitive Awareness Through Reconfiguration

This phase demonstrates metacognitive awareness through two reconfiguration scenarios: component deletion and hierarchical restructuring.

**Scenario A: Component Deletion**

**Input:** We delete the rounded-corner rectangle FSM from graph memory.

**Reconfiguration Process:** The system detects broken compositional dependencies by traversing edges from the deleted node. It identifies house HOAs containing compositional edges to the deleted rectangle FSM and marks them as structurally incomplete. The system then searches the rectangle cluster for similar alternatives to the deleted FSM, comparing the remaining rectangle FSMs against the deleted FSM's constraint structure using similarity detection. The remaining FSMs exceed the similarity threshold, making them viable substitutes. The affected house HOAs reconnect their compositional edges to available alternatives, completing the restructuring and maintaining structural coherence. If no rectangle FSM in the cluster had exceeded the similarity threshold, the affected house HOAs would have dissolved instead.

**Graph Evolution and Measurements:** The graph modification log records broken dependency edges, similarity-based search operations, edge reconnections, and restored structural coherence. This provides measurable metacognitive signatures: dependency awareness (affected structures identified), knowledge-based substitution (cluster search for alternatives), adaptive recovery (successful restructuring), and conditional dissolution (HOAs dissolve only when no similar alternatives exist). The system's response demonstrates that metacognitive awareness extends beyond detection to include knowledge-based repair strategies.

**Scenario B: Hierarchical Restructuring**

**Input:** We introduce line segments as more fundamental primitives. The system observes two distinct line patterns: horizontal lines and vertical lines.

**Reconfiguration Process:** The learning engine creates two line FSMs (horizontal and vertical). The system then analyzes existing rectangle and triangle FSMs to determine if they can be decomposed into these new primitives. Each rectangle matches a pattern of 4 lines (2 horizontal + 2 vertical forming closure). Each triangle matches a pattern of 3 lines forming closure. The system restructures all 3 rectangle FSMs into rectangle HOAs, each containing compositional edges to line FSMs. Similarly, all 3 triangle FSMs restructure into triangle HOAs. Consequently, the 4 house HOAs, which previously referenced rectangle FSMs and triangle FSMs, now reference rectangle HOAs and triangle HOAs, making houses 2nd-level HOAs.

**Graph Evolution and Measurements**: The system adds two line FSMs. Analysis reveals that existing rectangle and triangle FSMs can decompose into line compositions. The system restructures 6 FSMs (3 rectangle + 3 triangle) into HOAs, each containing compositional edges to line FSMs. The house HOAs update their references to point to rectangle HOAs and triangle HOAs, making houses 2nd-level HOAs. Measurements confirm ontological restructuring: 6 FSM-to-HOA conversions, hierarchical depth increase from 1 to 2 levels, and preserved cluster topology despite representational change. This demonstrates metacognitive capacity to refine knowledge architecture when discovering deeper primitives.

**Table 2.** Summary of measurements across all four phases.

| Phase | Measurement | Metric | Value | Criterion |
|---|---|---|---|---|

| 1 | Representation formation | Single graph transaction | O(1) complexity | Sample efficiency |
|---|---|---|---|---|
| 1 | Constraint capture | Graph structure inspection | Complete constraint graph | Structural knowledge |
| 2 | Primitive clustering | Average path length (rectangles) | 1 (from ∞) | Generalization capacity |
| 2 | Primitive clustering | Average path length (triangles) | 1 (from ∞) | Generalization capacity |
| 2 | Concept robustness | Modularity (both clusters) | 0.92 | Generalization capacity |
| 3 | Compositional construction | Dependency count | 2 components per house | Compositional awareness |
| 3 | Hierarchical depth | HOA nesting level | 1 level | Compositional awareness |
| 3 | Composite clustering | Average path length (houses) | 1.3 | Generalization capacity |
| 4A | Component deletion | Affected HOA count | 4 structures | Metacognitive access |
| 4B | Hierarchical restructuring | Depth increase | 1 level → 2 levels | Metacognitive access |
| 4B | Ontological restructuring | Node type changes | 6 FSMs → HOAs | Metacognitive access |
| 4B | Propagation scope | Affected structures | 4 house HOAs updated | Metacognitive access |

## 6 Analysis: Understanding Signatures vs Alternatives

The experimental demonstration showed that hierarchical automata produce measurable understanding signatures through graph properties. This section analyzes how these measurements differ from alternative approaches.

**Representation Formation: Statistical vs Structural:** Our system differs in three measurable ways. First, representation formation requires one exposure. A single square pattern creates its automaton immediately through constraint inference. This one-shot capability must be distinguished from robust concept formation: one observation creates one automaton, while robust concepts emerge through multiple exposures as similarity detection clusters related automata into stable patterns. This mirrors human learning: we recognize new objects immediately from one example but develop robust understanding through repeated encounters with variations.

Second, the representation captures structural necessity rather than statistical correlation. The house HOA encodes "houses require walls and roofs in specific spatial configuration", not "houses correlate with walls (p=0.95) and roofs (p=0.92)". Inspecting the constraint graph reveals deterministic dependencies (must-have relationships), not probabilistic weights (likely-to-co-occur values).

Third, concept robustness emerges through explicit clustering rather than implicit weight adjustment. When our system encounters house variations, it creates new automata and connects them via similarity edges. The resulting cluster structure is visible and quantifiable. Statistical systems refine representations by adjusting weights across the entire network, producing a distributed representation where concept boundaries remain implicit.

**Semantic Grounding: Association vs Constraint:** Beyond representation formation, understanding requires that symbols acquire meaning through grounding. Association-based approaches ground symbols through co-occurrence. The symbol "house" acquires meaning by appearing in contexts with walls, windows, etc. Neural language models ground words through distributional semantics: words appearing in similar contexts acquire similar embeddings.

Our approach grounds symbols through structural constraints inferred from observation. When the learning engine constructs a house automaton, it analyzes the pattern to determine intrinsic structural properties: walls must be vertical, roofs must sit atop walls, spatial relationships must satisfy specific constraints. The measurement difference is representation type. Association-based grounding produces correlation matrices or embedding vectors we can inspect for similarity but not for structural relationships. Our constraint graphs show explicit dependencies. Following edges from house to walls reveals that houses structurally require walls as components. This affects how we measure understanding. Association-based systems test grounding through behavioral probes: does the model predict "wall" given "house" context? These are indirect measurements that infer understanding from performance. In our system, we directly inspect the constraint structure: query the house HOA, extract its compositional graph, verify wall dependencies with spatial constraints.

The grounding mechanism also affects sample efficiency. Association-based approaches require extensive exposure to build reliable correlation statistics because distributional patterns only emerge across many contexts. Our constraint inference operates from single observations because structural properties are intrinsic to patterns.

**Table 3.** Measurement Comparison

| Property | Statistical approaches | Our approach |
|---|---|---|
| Understanding formation | Gradual (loss curves, accuracy) | Discrete (graph modifications) |
| Compositional structure | Implicit (embedding geometry) | Explicit (HOA topology) |
| Concept robustness | Distributed (weight adjustments) | Structural (cluster metrics) |
| Grounding mechanism | Co-occurrence statistics | Constraint inference |
| Sample efficiency | Thousands of examples | One observation per automaton |
| Verification method | Behavioral probes (indirect) | Graph inspection (direct) |

# 7 Discussion

In this section we discuss implications for AGI research, connections to human cognition, requirements for trustworthy AI systems, and limitations that suggest future work.

## 7.1 Addressing the Measurement Gap

The measurement gap in current AI systems has practical consequences. We can measure performance but not understanding itself. Deploying systems in critical domains requires confidence that they understand task requirements, not just correlate inputs with outputs. Current approaches verify understanding indirectly through extensive testing. But adversarial examples demonstrate that high performance does not guarantee robust understanding.

Hierarchical automata address this gap by making understanding directly inspectable. When we ask "does this system understand houses?", we query the graph for house automata and extract compositional structure, constraints, and clustering patterns. The answer is read from structure, not inferred from performance. This inspection provides a verification mechanism: the graph reveals what concepts exist, how they compose, and how robust they are to variation, distinguishing comprehension from correlation.

This verification capability complements perceptual capabilities in neural systems. Neural approaches excel at learning from raw sensory data. They extract features from pixels, learn from massive datasets, and handle continuous variation. We make compositional knowledge formation measurable through discrete structures. Complete AI systems will likely integrate both capabilities. Neural perception produces symbolic observations. Our discrete structures make understanding formation measurable. Neuro-symbolic execution uses discovered compositional knowledge for task performance.

## 7.2 Sample Efficiency and Human-Like Learning

Human learning demonstrates remarkable sample efficiency. We see one example of a novel object category and can recognize other instances despite significant variation. Fast mapping shows that word meanings are acquired from single exposures and refined through subsequent experience.

The immediate representation mechanism in our system mimics this pattern. Single exposure creates an automaton through constraint inference that captures structural properties but lacks robustness to variation. Subsequent exposures create additional automata with similarity edges, and the resulting cluster structure provides robust concept coverage. This two-stage process matches human concept development more closely than statistical accumulation or practice-based compilation.

Human concepts organize around prototype-like structures with graded membership. Our cluster structure exhibits the same property: central automata have high connection degree and strong similarity edges, peripheral automata have fewer or weaker connections. Average path length from a query automaton to cluster members predicts how "typical" an instance is, matching human judgments about category membership.

### 7.3 Compositional Transparency and Explainable AI

Explainable AI addresses the challenge of making system reasoning interpretable to humans. Neural networks pose particular challenges: their decisions emerge from millions of parameter interactions that resist comprehension. Post-hoc explanation methods such as LIME (Ribeiro et al., 2016) and SHAP (Lundberg & Lee, 2017) provide partial insight into decisions but cannot fully expose the reasoning process.

Hierarchical automata provide transparency through compositional structure. When the system identifies a pattern as a house, we explain this by exposing the compositional graph: the pattern contains wall and roof components connected by spatial relationships the house concept requires. Rather than explaining "the network activated strongly for house-like features," we state "the pattern satisfies the structural constraints that define houses." Such transparency extends to metacognitive reasoning. When we delete a component or add deeper primitives and observe reconfiguration, we explain the system's response by showing broken dependency edges and affected structures. The explanation reveals the system's knowledge architecture showing which concepts depended on the deleted component and how the network reorganized to maintain consistency. Compositional transparency is important for trustworthy AI deployment. In domains where decisions have consequences, we need to verify that systems base decisions on appropriate factors and understand relevant constraints. Graph-based representations make verification possible: we can inspect the knowledge structure to confirm relevant dependencies exist and inappropriate biases are absent.

### 7.4 Limitations and Future Directions

The current work demonstrates measurable understanding in a simple geometric domain. Four significant limitations suggest directions for future research.

**Domain Complexity**: Geometric primitives represent a restricted domain with clear structural constraints and unambiguous composition rules. Scaling to complex domains introduces challenges: constraints become less crisp, composition becomes less hierarchical as concepts interconnect in networks, and variation becomes harder to characterize across multiple dimensions. Additionally, the current implementation assumes perceptual primitives are pre-defined: geometric constraints such as segment length, angle, and spatial position are given rather than learned from raw perception. Grounding these structural representations at the perceptual level (the problem Harnad (1990) identified) remains an open challenge separate from the compositional learning demonstrated here.

**Continuous Learning**: The demonstration shows discrete learning episodes. Real-world learning is often continuous: new patterns arrive constantly, concepts evolve gradually, and the system must integrate new knowledge without catastrophic forgetting. The graph structure supports continuous addition without disrupting existing structure, but questions remain about long-term memory management, concept drift, and maintaining coherence as the graph grows.

**Integration with Other Mechanisms**: The current architecture focuses on structural learning from observation. It does not address other aspects of intelligence: planning requires temporal reasoning, language requires syntax and pragmatics, social cognition requires theory of mind. A complete AGI system needs multiple mechanisms working together.

**Scalability**: The demonstration involves small graphs (tens of nodes). Real-world knowledge bases contain thousands or millions of concepts. Graph operations must scale efficiently. Questions remain about computational complexity as graphs grow large.

These limitations do not undermine the central contribution: hierarchical automata make understanding measurable through structural properties. Applying this framework to full AGI systems will require extending the approach to handle complexity, continuous learning, mechanism integration, and scale.

So, "how do we measure genuine understanding?" This work proposes an answer: build systems where understanding formation produces discrete structural signatures we can inspect, compositional knowledge is explicitly represented in hierarchical graphs, concept robustness is quantifiable through topology metrics, and metacognitive awareness is observable through reconfiguration behavior. This differs from current approaches that measure understanding indirectly through performance or estimate it through confidence scores. Direct measurement requires representational substrates that make understanding structural rather than statistical, discrete rather than continuous, and inspectable rather than distributed.

The path from measuring understanding in geometric domains to measuring understanding in full AGI systems is not trivial. But our framework establishes that understanding can be measurable if we design architectures specifically to produce measurable signatures. This suggests a new research direction in which we ask "how do we build AI systems where understanding is inherently measurable?" instead of "how do we measure understanding in existing AI systems?

**Disclosure of Interests.** The authors have no competing interests to declare that are relevant to the content of this article.

## References

1. Anderson, J.R.: How Can the Human Mind Occur in the Physical Universe? Oxford University Press, New York (2007)
2. Anderson, J.R., Lebiere, C.: The Atomic Components of Thought. Lawrence Erlbaum Associates, Mahwah (1998)
3. Andreas, J., Rohrbach, M., Darrell, T., Klein, D.: Neural module networks. In: Proceedings of the IEEE Conference on Computer Vision and Pattern Recognition, pp. 39-48 (2016)
4. Carey, S., Bartlett, E.: Acquiring a single new word. Papers and Reports on Child Language Development 15, 17-29 (1978)
5. Elhage, N., Nanda, N., Olsson, C., Henighan, T., Joseph, N., Mann, B., Askell, A., Bai, Y., Chen, A., Conerly, T., DasSarma, N., Drain, D., Ganguli, D., Hatfield-Dodds, Z., Hernandez, D., Jones, A., Kernion, J., Lovitt, L., Ndousse, K., Amodei, D., Brown, T., Clark, J., Kaplan, J., McCandlish, S., Olah, C.: A Mathematical Framework for Transformer Circuits. Transformer Circuits Thread, https://transformer-circuits.pub/2021/framework/index.html (2021) last accessed 2026/05/30

6. Finn, C., Abbeel, P., Levine, S.: Model-agnostic meta-learning for fast adaptation of deep networks. In: Proceedings of the 34th International Conference on Machine Learning, vol. 70, pp. 1126-1135, JMLR, Sydney (2017)
7. Flavell, J.H.: Metacognition and cognitive monitoring: A new area of cognitive-developmental inquiry. American Psychologist 34(10), 906-911 (1979)
8. Fodor, J.A.: The Language of Thought. Harvard University Press, Cambridge (1975)
9. Goertzel, B., Lian, R., Arel, I., de Garis, H., Chen, S.: OpenCog: A software framework for integrative Artificial General Intelligence. In: Goertzel, B., Pennachin, L., Geisweiller, N. (eds.) Engineering General Intelligence, Part 1: A Path to Advanced AGI via Embodied Learning and Cognitive Synergy, pp. 3-29. Atlantis Press, Dordrecht (2014)
10. Goertzel, B., Looks, M., Pennachin, C., de Garis, H.: Probabilistic Logic Networks. In: Goertzel, B., Hitzler, P., Hutter, M. (eds.) Artificial General Intelligence 2008: Proceedings of the First AGI Conference, pp. 178-189. IOS Press, Amsterdam (2008)
11. Harnad, S.: The symbol grounding problem. Physica D: Nonlinear Phenomena 42(1-3), 335–346 (1990)
12. Laird, J.E.: The Soar Cognitive Architecture. MIT Press, Cambridge (2012)
13. Laird, J.E., Newell, A., Rosenbloom, P.S.: Soar: An architecture for general intelligence. Artificial Intelligence 33(1), 1-64 (1987)
14. Li, Z., Chen, J., Huang, K., Wu, W., Zhang, C., Huang, Z., Wang, W.Y.: Scallop: A language for neurosymbolic programming. ACM Program. Lang. 7, PLDI, Article 166 (2023)
15. Lundberg, S.M., Lee, S.I.: A unified approach to interpreting model predictions. In: NIPS'17: Proceedings of the 31st International Conference on Neural Information Processing Systems. pp 4768-4777. Curran Associates Inc., NY, USA (2017)
16. Mandler, J.M.: The Foundations of Mind: Origins of Conceptual Thought. Oxford University Press, New York (2004)
17. Manhaeve, R., Dumancic, S., Kimmig, A., Demeester, T., De Raedt, L.: Neural probabilistic logic programming in DeepProbLog. Artificial Intelligence, vol. 298, 103504 (2021)
18. Ribeiro, M.T., Singh, S., Guestrin, C.: "Why should I trust you?": Explaining the predictions of any classifier. In: Proceedings of the 22nd ACM SIGKDD International Conference on Knowledge Discovery and Data Mining, pp. 1135-1144. Association for Computing Machinery, New York (2016)
19. Rosch, E.: Principles of categorization. In: Rosch, E., Lloyd, B.B. (eds.) Cognition and Categorization, pp. 27-48. Lawrence Erlbaum Associates, Hillsdale (1978)
20. Rosenbloom, P.S.: On Computing: The Fourth Great Scientific Domain. MIT Press, Cambridge (2013)
21. Rosenbloom, P.S., Demski, A., Ustun, V.: Extending Cognitive Architectures with Mental Imagery. In: Langley, P. (ed.) Proceedings of the Second Annual Conference on Advances in Cognitive Systems, pp. 77-94. Cognitive Systems Foundation (2013)
22. Searle, J.R.: Minds, brains, and programs. Behavioral and Brain Sciences 3(3), 417-424 (1980)
23. Snell, J., Swersky, K., Zemel, R.: Prototypical networks for few-shot learning. In: NIPS'17: Proceedings of the 31st International Conference on Neural Information Processing Systems
24. Pages 4080 – 4090. Curran Associates Inc., NY (2017)
25. Sun, R.: Anatomy of the Mind: Exploring Psychological Mechanisms and Processes with the Clarion Cognitive Architecture. Oxford University Press, New York (2016)
26. Sun, R., Merrill, E., Peterson, T.: From implicit skills to explicit knowledge: A bottom-up model of skill learning. Cognitive Science 25(2), 203-244 (2001)

27. Vaswani, A., Shazeer, N., Parmar, N., Uszkoreit, J., Jones, L., Gomez, A.N., Kaiser, Ł., Polosukhin, I.: Attention is all you need. In: Advances in Neural Information Processing Systems (NeurIPS 2017), pp. 5998-6008 (2017)